\documentclass[onecolumn,showpacs]{revtex4}

\usepackage{graphicx}% Include figure files
\usepackage{dcolumn}% Align table columns on decimal point
\usepackage{bm}% bold math

\input epsf

\begin{document}

\title{\Large Interaction between DBI-essence and other Dark Energies }

\author{\bf  Surajit
Chattopadhyay$^1$\footnote{surajit$_{_{-}}2008$@yahoo.co.in} and
Ujjal Debnath$^2$\footnote{ujjaldebnath@yahoo.com ,
ujjal@iucaa.ernet.in}}

\affiliation{$^1$Department of Computer Application, Pailan
College of Management and Technology, Bengal Pailan Park,
Kolkata-700 104, India.\\
$^2$Department of Mathematics, Bengal Engineering and Science
University, Shibpur, Howrah-711 103, India. }

\date{\today}

\begin{abstract}
The present work considers interaction between DBI-essence and
other candidates of dark energies like modified Chaplygin gas,
hessence, tachyonic field, and new agegraphic dark energy. The
potentials of the fields have been reconstructed under interaction
and their evolutions have been viewed against cosmic time $t$ and
scalar field $\phi$. Equation of state parameters have also been
obtained. The nature of potentials and the equation of state
parameters of the dark energies have been found graphically in
presence of interaction (both small and large interaction).
\end{abstract}


\maketitle

\section{\normalsize\bf{Introduction}}

Plethora of literatures have discussed the fact that the universe
is not only expanding, but it is accelerating also [1,2]. Since
the discovery of the fact that presently there is an accelerating
expansion of the universe; different suggestions have been made in
order to explain the accelerated expansion. Among the various
suggestions, one is the hypothesis of dark energy, which is a
fluid of negative pressure representing about $70\%$ of the total
energy of the universe [3]. The dark energy problem continues to
be a sticking point for theoretical physicists. The simplest
solution to this problem is to postulate the existence of a vacuum
energy or cosmological constant $\Lambda$, which agrees with all
the current observational bounds [4]. The cosmological constant
corresponds to a fluid with a constant equation of state
$\omega=-1$. Now, the observations which constrain the value of
$\omega$ today to be close to that of the cosmological constant
actually say relatively little about the time evolution of
$\omega$, and so we can broaden our horizons and consider a
situation in which the equation of state of dark energy changes
with time, such as in inflationary cosmology. In 1998, the
accelerated expansion of the universe was pointed out by two
groups from the observations of Type Ia Supernova (SNe Ia).
Unified models of the two main components of the universe, dark
energy and dark matter, represent an interesting option for
explaining the substantial evidence of the current acceleration of
the universe. Following the tradition of trying to find an
interconnection between the world of Particle Physics and
Cosmology, it is customary to try and view unified dark energy
models as scalar field scenarios with much steeper potentials than
in the standard weakly coupled slow roll inflation model
[5].\\

There have been many works aimed at connecting the string theory
with inflation. While doing so, various ideas in string theory
based on the concept of branes have proved themselves fruitful.
Scenarios where the inflation is interpreted as the distance
between two branes moving in the extra dimension along a warped
throat have given rise to many interesting studies [6]. One area
which has been well explored in recent years, is inflation driven
by the open string sector through dynamical Dp-branes. This is the
so-called DBI (Dirac-Born-Infield) inflation [7], which lies in a
special class of K-inflation models. It was originally thought
that such models yielded large levels of non-Gaussian
perturbations which could be used as a falsifiable signature of
string theory [3,7]. However subsequent work has shown that this
is may not be the case, and that the simplest DBI models are
essentially indistinguishable from standard field theoretic slow
roll models [8]. Martin and Yamaguchi [3] introduced a scalar
field model where the kinetic term has a DBI form and considered
that the dark energy scalar field is a DBI scalar field, for which
the action of the field can be written as

\begin{equation}
S_{DBI}=-\int d^{4}xa^{3}
(t)\left[T(\phi)\sqrt{1-\frac{\dot{\phi}^{2}}{T(\phi)}}+V(\phi)-T(\phi)\right]
\end{equation}

where $T(\phi)$ is the tension and $V(\phi)$ is the potential. \\

To obtain a suitable evolution of the Universe an interaction is
often assumed such that the decay rate should be proportional to
the present value of the Hubble parameter for good fit to the
expansion history of the Universe as determined by the Supernovae
and CMB data [9]. These kind of models describe an energy flow
between the components so that no components are conserved
separately. There are several work on the interaction between dark
energy (tachyon or phantom) and dark matter [10], where
phenomenologically introduced different forms of interaction term.
\\

In the present work, we would consider the interactions between
different candidates of dark energy and DBI-essence model. As
other models of dark energy we would consider Chaplygin gas,
hessence and tachyonic field. Organization of the rest of the
paper is as follows. In section II, we would consider the basic
equations. In section III, we would consider the interaction
between DBI-essence and Chaplygin gas and in section IV, we would
consider the interaction between DBI-essence and hessence. The
interaction between DBI-essence and tachyonic field would be
considered in section V. In section VI, the interaction between
DBI-essence and new agegraphic dark energy would be considered.
The results would be discussed in section VII.
 \\

\section{\normalsize\bf{Basic Equations}}

We consider a spatially flat isotropic and homogeneous universe in
the FRW model whose metric is given by

\begin{equation}
ds^{2}=dt^{2}-a^{2}(t)[dr^{2}+r^{2}(d\theta^{2}+sin^{2}d\phi^{2})]
\end{equation}
where, $a(t)$ is the scale factor. The Einstein field equations
are given by (choosing $8\pi G=c=1$)

\begin{equation}
3\frac{\dot{a}^{2}}{a^{2}}=\rho
\end{equation}

\begin{equation}
6\frac{\ddot{a}}{a}=-(\rho+3 p)
\end{equation}

The energy conservation equation is given by

\begin{equation}
\dot{\rho}+3\frac{\dot{a}}{a} (\rho+p)=0
\end{equation}

If we consider a model consisting of two component mixture, the an
interaction term needs to be introduced. In a two-component model,
we replace $\rho$ and $p$ of equations (3), (4) and (5) by
$\rho_{total}$ and $p_{total}$ where

\begin{equation}
\rho_{total}=\rho_{D}+\rho_{X}
\end{equation}

\begin{equation}
p_{total}=p_{D}+p_{X}
\end{equation}

where $\rho_{D}$ and $p_{D}$ denote the density and pressure for
the DBI-essence. The terms $\rho_{X}$ and $p_{X}$ denote the
density and pressure corresponding to the other dark energies.
Therefore, equations (3), (4) and (5) get modified to

\begin{equation}
3\frac{\dot{a}^{2}}{a^{2}}=(\rho_{D}+\rho_{X})
\end{equation}

\begin{equation}
6\frac{\ddot{a}}{a}=-\left[(\rho_{D}+\rho_{X})+3(p_{D}+p_{X})\right]
\end{equation}

\begin{equation}
(\dot{\rho}_{D}+\dot{\rho_{X}})+3\frac{\dot{a}}{a}[(\rho_{D}+\rho_{X})+(p_{D}+p_{X})]=0
\end{equation}

Assuming gravity to obey four-dimensional general relativity with
a standard Einstein-Hilbert Lagrangian, the density and pressure
for DBI-essence are read as [7]

\begin{equation}
\rho_{D}=(\gamma-1)T(\phi_{D})+V_{D}(\phi_{D})
\end{equation}

\begin{equation}
p_{D}=\left(\frac{\gamma-1}{\gamma}\right)T(\phi_{D})-V_{D}(\phi_{D})
\end{equation}

where, $T(\phi_{D})$ is the tension, $V_{D}(\phi_{D})$ is the
potential, $\phi_{D}$ denotes the scalar field for DBI-essence and
the quantity $\gamma$ is reminiscent of the usual Lorentz factor
given by

\begin{equation}
\gamma=\frac{1}{\sqrt{1-\frac{\dot{\phi}_{D}^{2}}{T(\phi_{D})}}}
\end{equation}

Since we are considering two-component model, we consider the
interaction term $3H\delta\rho_{X}$ and using (10) we can write
the conservation equations as

\begin{equation}
\dot{\rho}_{D}+3H(\rho_{D}+p_{D})=3H\delta\rho_{X}
\end{equation}

\begin{equation}
\dot{\rho}_{X}+3H(\rho_{X}+p_{X})=-3H\delta\rho_{X}
\end{equation}

where, $H=\frac{\dot{a}}{a}$ is the Hubble parameter, $\delta$ is
the interaction parameter and rest of the symbols are as explained
earlier.\\

\section{\normalsize\bf{Interaction with Chaplygin gas}}

Recently the so-called Chaplygin gas, also dubbed quartessence,
was suggested as a candidate of a unified model of dark energy and
dark matter [11]. Pure Chaplygin gas is a candidate for dark
energy, which obeys an exotic equation of state

\begin{equation}
p_{ch}=-\frac{A}{\rho_{ch}}
\end{equation}

where, where $p_{ch}$ and $\rho_{ch}$ are  the pressure and energy
density respectively and $A$ is a positive constant. This equation
leads to a density evolution in the form

\begin{equation}
\rho_{ch}=\sqrt{A+\frac{B}{a^{6}}}
\end{equation}

where, $B$ is an integration constant. The model unifies both dark
energy and dark matter. The reason is that, from (17), the
Chaplygin gas behaves as dust-like matter at early stage and as a
cosmological constant at later stage. The Chaplygin gas emerges as
an effective fluid associated with D-branes and can also be
obtained from the Born-Infeld action [11]. Recently, the original
Chaplygin gas model was generalized, and  subsequently the above
equation (16) was modified to the form (known as \emph{generalized
Chaplygin gas})[12]

\begin{equation}
p_{ch}=-\frac{A}{\rho_{ch}^{\alpha}}~~~~,~~~~~~0\leq \alpha \leq 1
\end{equation}

This generalized Chaplygin gas (GCG) model has been studied
previously [12]. There are some works on modified Chaplygin gas
(MCG) obeying an equation of state [11]

\begin{equation}
p_{ch}=A\rho_{ch}-\frac{B}{\rho_{ch}^\alpha}
\end{equation}

and the corresponding evolution of density is

\begin{equation}
\rho_{ch}=\left[\frac{B}{1+A}+\frac{C}{a^{3(1+\alpha)(1+A)}}\right]^{\frac{1}{1+\alpha}}
\end{equation}

where, $A$, $B$ and $C$ are positive constants. It is obvious from
the equation (18) that if $\alpha=1$, then GCG model becomes the
original Chaplygin gas model. It can be seen very easily that this
energy density interpolates between a dust-like configuration
$(\rho\simeq \sqrt{B}a^{-3},~p\simeq 0)$ in the past and a
de-Sitter-like one $(p=-\rho)$ in the late times. This property
makes the GCG model an interesting candidate for the unification
of dark matter and dark energy. Zhang and Zhu (Ref. [11])
considered an interaction between ordinary Chaplygin gas model and
dark matter. Ref [13] has shown that the expression for $H$
obtained from GCG is a good fit for the observational data as far
as the background cosmology is
concerned.\\

 Since our purpose is to consider interaction between
DBI-essence and MCG obeying the equation of state given by
equation (19); we consider the conservation equations (14) and
(15) involving the interaction term $\delta$. In the present case
$\rho_{X}\equiv\rho_{ch}$ and $p_{X}\equiv p_{ch}$. We choose
 $T=n\dot{\phi}_{D}^{2}$ and $a=t^{m}$ in equations (8)
and (11)-(15). We have chosen $T=n\dot{\phi}_{D}^{2}$ because the
term $\dot{\phi}_{D}^{2}$ is already there in the expression for
$\gamma$ and consequently this choice makes the calculation
convenient. Since it is already mentioned that the tension $T$ is
a function of the field $\phi_{D}$, the choice seems suitable.
Since scale factor $a$ is a function of time $t$, we have chosen
it in the power law form as $t^{m}$ with $m>0$ and since it
matches the accelerated expansion of the universe, this choice
seems suitable. For such choices, we get under interaction, the
expression for the density of Chaplygin gas as

\begin{equation}
\rho_{ch}=\left(\frac{B}{1+A+\delta}+\frac{C}{t^{3m(1+\alpha)(1+A+\delta)}}\right)^{\frac{1}{1+\alpha}}
\end{equation}

Also the expressions of scalar field and the relevant potential
for DBI-essence under interaction are obtained as

\begin{eqnarray*}
\phi_{D}=\int \left[2\sqrt{\frac{n-1}{n}}
\left\{\frac{m}{t^{2}}+\frac{1}{2}\left(\frac{B}{1+A+\delta}+\frac{C}{t^{3m(1+\alpha)(1+A+\delta)}}\right)^{-\frac{\alpha}{1+\alpha}}\times\right.\right.
\end{eqnarray*}

\begin{equation}
\left.\left.\left(B-(1+A+\delta)\left(\frac{B}{1+A+\delta}+\frac{C}{t^{3m(1+\alpha)(1+A+\delta)}}\right)\right)\right\}\right]^{1/2}dt
\end{equation}
and
\begin{equation}
V_{D}=\frac{3m^{2}}{t^{2}}-(\gamma-1)n\dot{\phi}_{D}^{2}-\rho_{ch}
\end{equation}

Figures 1 and 2 show the variation of $\phi_{D}$ against $t$ and
the variation of $V_{D}$ against $\phi_{D}$ respectively in
presence of interaction between DBI-essence and MCG. From these
figures we see that the DBI scalar field increases with cosmic
time $t$, whereas the DBI potential decreases with the field under
the interaction.\\

\begin{figure}

\includegraphics[height=1.8in]{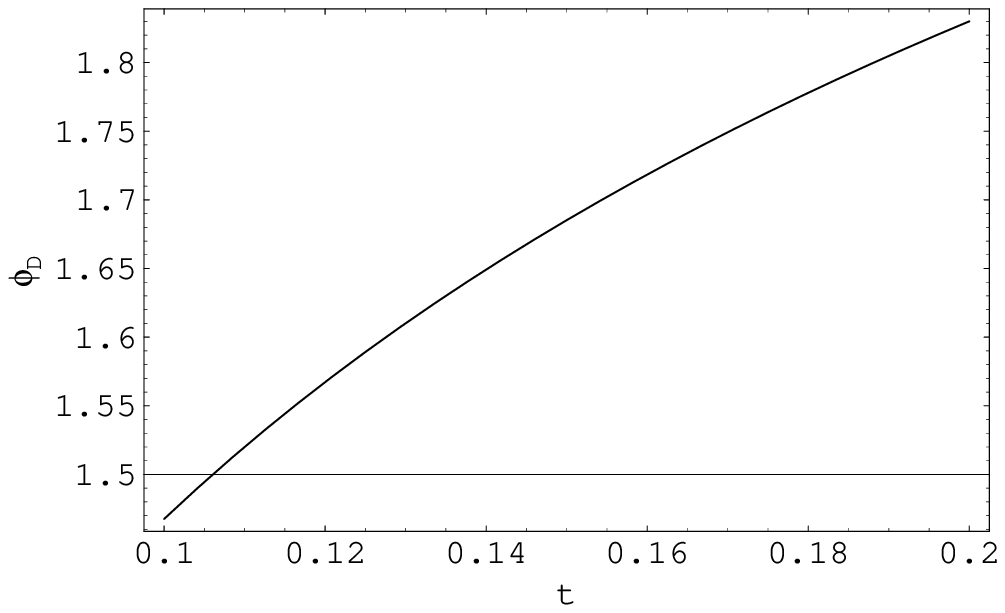}~~~
\includegraphics[height=1.8in]{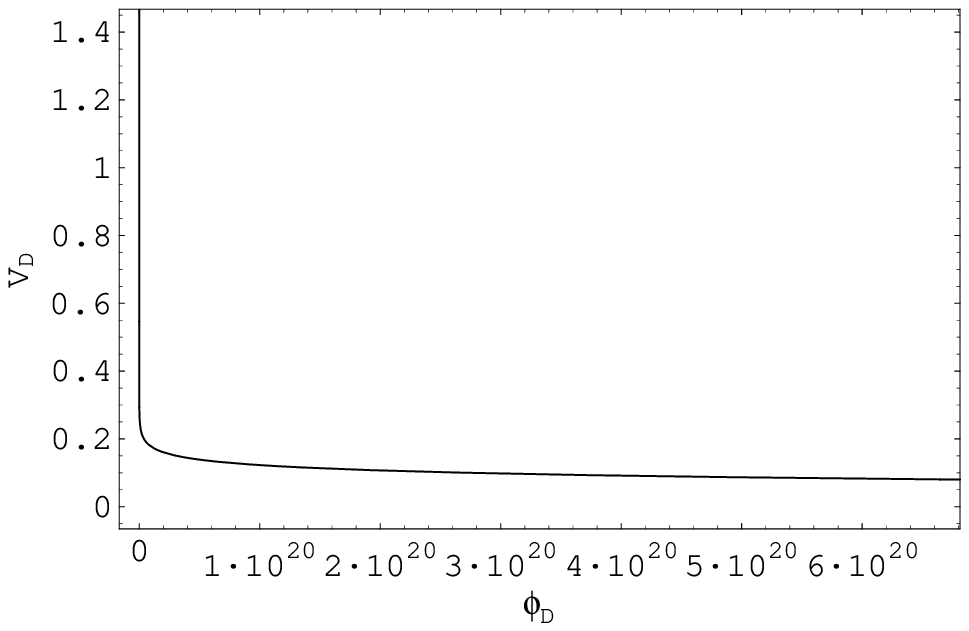}~\\
\vspace{1mm} ~~~~~~~~~~~~Fig.1~~~~~~~~~~~~~~~~~~~~~~~~~~~~~~~~~~~~~~~~~~~~~~~~~~~~~~~~Fig.2\\
\vspace{6mm} Figs. 1 and 2 show the variation of $\phi_{D}$
against $t$ and the variation of $V_{D}$ against $\phi_{D}$
respectively in presence of interaction between DBI-essence and
MCG.

\vspace{6mm} \vspace{6mm}

\end{figure}

\section{\normalsize\bf{Interaction with Hessence}}

The hessence is a kind of simple quintom, which has the Lagrangian
density [14]

\begin{equation}
{\cal
L}_{DE}=\frac{1}{2}[(\partial_{\mu}\phi_{1})^{2}-(\partial_{\mu}\phi_{2})^{2}]-V(\phi_{1},\phi_{2})
\end{equation}

where $\phi_{1}$ and $\phi_{2}$ are two real scalar fields and
play the roles of quintessence and phantom respectively.
Considering a spatially flat Friedmann-Robertson-Walker (FRW)
universe and assuming the scalar fields $\phi_{1}$ and $\phi_{2}$
are homogeneous, one obtains the effective equation of state as

\begin{equation}
\omega=\frac{\dot{\phi}_{1}^{2}-\dot{\phi}_{2}^{2}-2V(\phi_{1},\phi_{2})}{\dot{\phi}_{1}^{2}-\dot{\phi}_{2}^{2}+2V(\phi_{1},\phi_{2})}
\end{equation}

It is obvious that for $\dot{\phi}_{1}^{2}>\dot{\phi}_{2}^{2}$ we
get $\omega>-1$ and for $\dot{\phi}_{1}^{2}<\dot{\phi}_{2}^{2}$ we
get $\omega<-1$. In the present work, we consider a quintom model
having $V(\phi_{1},\phi_{2})=V(\phi_{1}^{2}-\phi_{2}^{2})$.
Similar choice of $V$ is available in Wei et al of reference [14].
We consider the action

\begin{equation}
S=\int d^{4}x\sqrt{-g}\left(-\frac{R}{16\pi G}+{\cal L}_{DE}+{\cal
L}_{m}\right)
\end{equation}

where, $g$ is the determinant of the metric $g_{\mu\nu}$, $R$ is
the Ricci scalar, ${\cal L}_{DE}$ and ${\cal L}_{m}$ are the
Lagrangian densities of dark energy and dark matter respectively.
The Lagrangian density of Hessence is given by the equation (23).
It can be shown that the Lagrangian remains invariant under the
transformation

\begin{equation}
\begin{array}{l}
\phi_{1}\rightarrow\phi_{1}\cos\alpha-i\phi_{2}\sin\alpha\\\\
\phi_{2}\rightarrow-i\phi_{1}\sin\alpha+\phi_{2}\cos\alpha
\end{array}
\end{equation}

\begin{figure}
\includegraphics[height=1.5in]{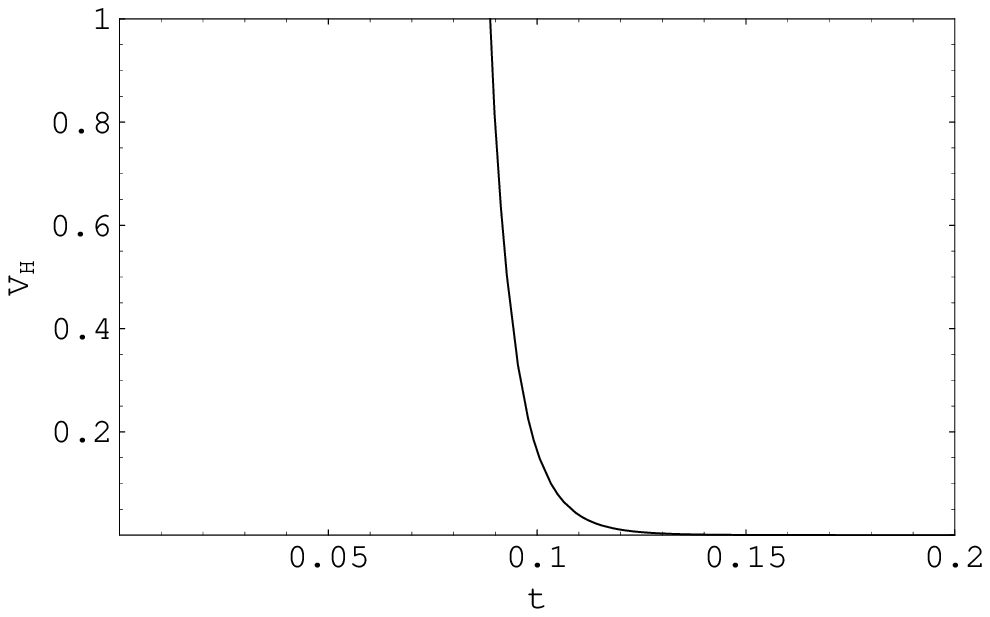}~~~~
\includegraphics[height=1.5in]{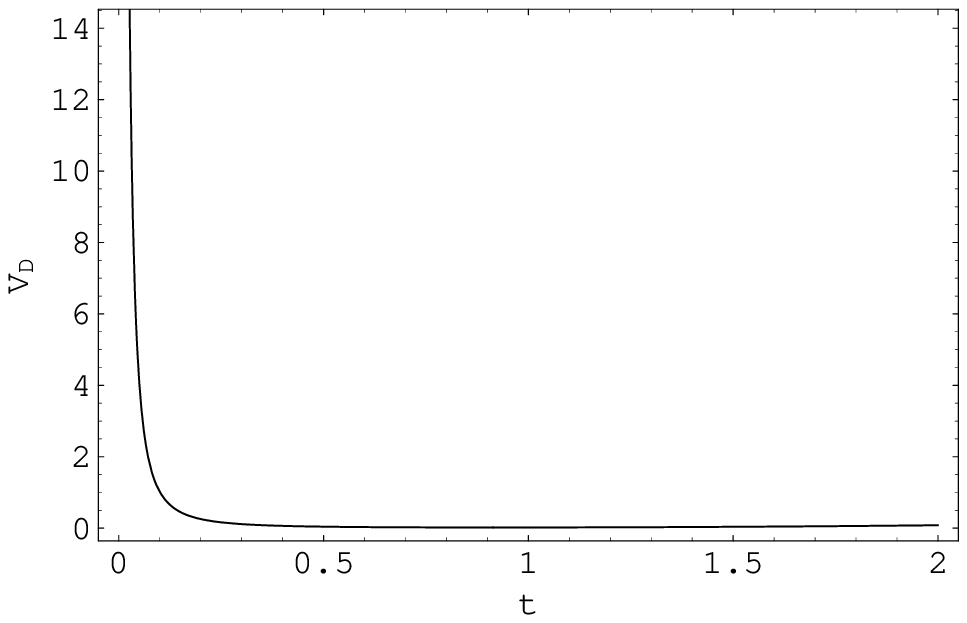}~\\
\vspace{1mm} ~~~~~~~~~~~~Fig.3~~~~~~~~~~~~~~~~~~~~~~~~~~~~~~~~~~~~~~~~~~~~~~~~~Fig.4\\
\includegraphics[height=1.5in]{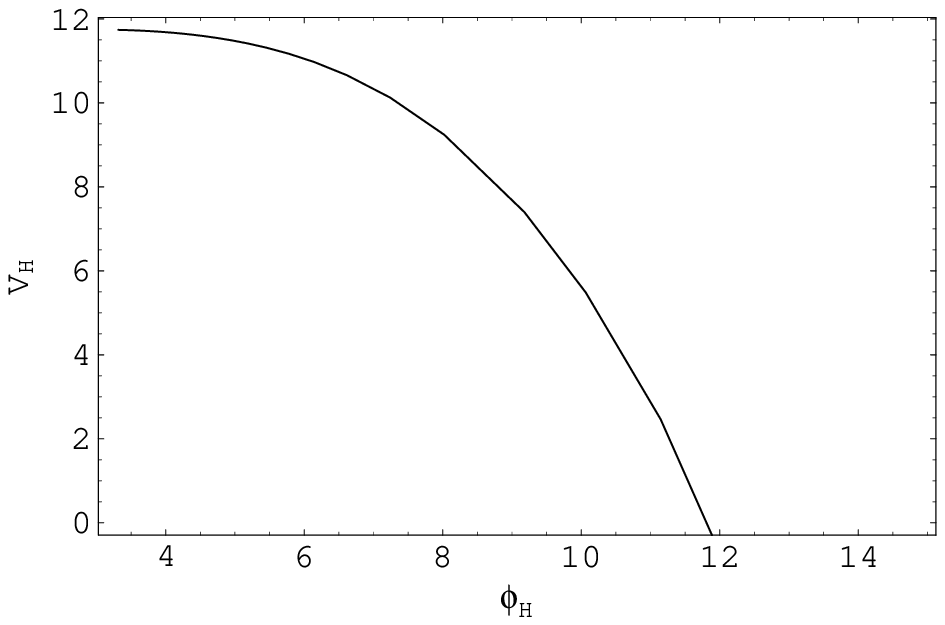}~~~~
\includegraphics[height=1.5in]{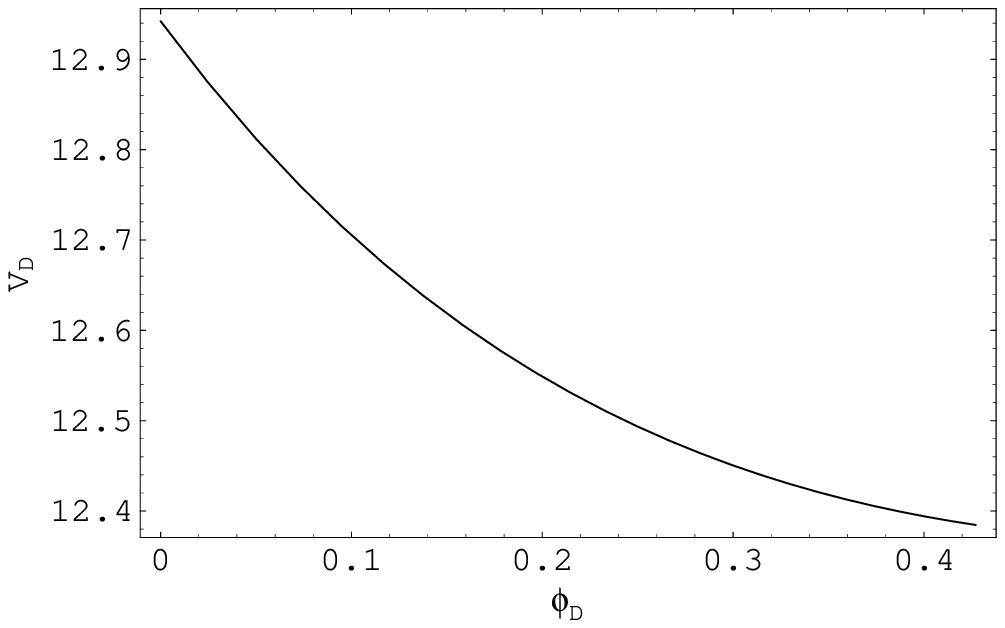}~\\
\vspace{1mm} ~~~~~~~~~~~~Fig.5~~~~~~~~~~~~~~~~~~~~~~~~~~~~~~~~~~~~~~~~~~~~~~~~~Fig.6\\
\includegraphics[height=1.5in]{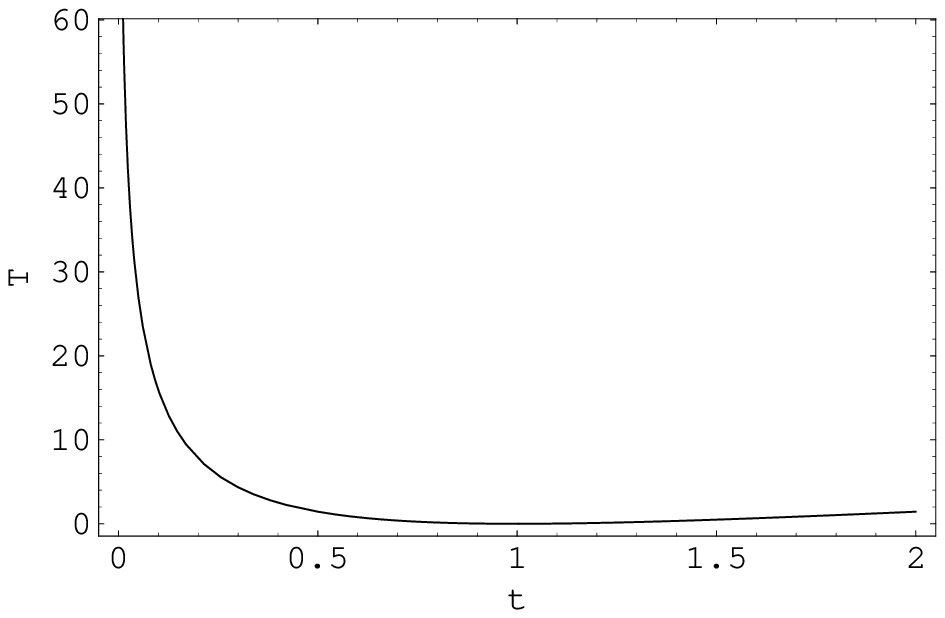}~\\
\vspace{1mm} ~~~~~~~~~~~~Fig.7\\
\vspace{6mm} Figs. 3 and 4 show the variation of $V_{H}$ and
$V_{D}$against $t$ in the case of interaction between DBI-essence
and hessence.\\ Figs. 5 and 6 show the variations of the above
potentials with the corresponding scalar fields in the presence of
the said interaction.\\
Fig. 7 shows the variation of $T(\phi_{D})$ with time $t$ in the
presence of the said interaction.

\vspace{6mm} \vspace{6mm}
\end{figure}

Consequently, the Lagrangian density equation (24) gets the form

\begin{equation}
{\cal
L}_{DE}=\frac{1}{2}[(\partial_{\mu}\phi)^{2}-\phi^{2}(\partial_{\mu}\theta)^{2}]-V(\phi)
\end{equation}

where, the new variables $(\phi, \theta)$ are given as

\begin{equation}
\begin{array}{l}
\phi_{1}=\phi \cosh \theta~~~;~~~~~~~~ \phi_{2}=\phi \sinh \theta
\end{array}
\end{equation}

From the action given in (26) we get under the assumption of
homogeneous $\phi$ and $\theta$ it can be obtained [14]

\begin{equation}
\ddot{\phi}+3H \dot{\phi}+\phi \dot{\theta}^{2}+dV/d\phi=0
\end{equation}

\begin{equation}
\phi^{2}\ddot{\theta}+(2\phi\dot{\phi}+3H\phi^{2})\dot{\theta}=0
\end{equation}

Detailed derivation of the above equations (30) and (31) is
available in Wei et al of reference [14].\\

Equation (31) implies that

\begin{equation}
Q=a^{3}\phi^{2}\dot{\theta}=Constant
\end{equation}

which is associated with the total conserved charge within the
physical volume [15]. This relation gives

\begin{equation}
\dot{\theta}=\frac{Q}{a^{3}\phi^{2}}
\end{equation}

From equation (30) one finds that the sign of the conserved charge
Q is determined by the sign of $\dot{\theta}$. The conserved
charge Q is positive for the case $\dot{\theta}> 0$ while Q is
negative for the case of $\dot{\theta}< 0$.

From equations (30) and (33) we get

\begin{equation}
\ddot{\phi}+3H\dot{\phi}+\frac{Q^{2}}{a^{6}\phi^{3}}+\frac{dV}{d\phi}=0
\end{equation}

Denoting the scalar field $\phi$ of the hessence by $\phi_{H}$ we
get the pressure and density of the hessence as

\begin{equation}
\rho_{H}=\frac{1}{2}(\dot{\phi}_{H}^{2}-\phi_{H}^{2}\dot{\theta}^{2})+V_{H}(\phi_{H})
\end{equation}

\begin{equation}
p_{H}=\frac{1}{2}(\dot{\phi}_{H}^{2}-\phi_{H}^{2}\dot{\theta}^{2})-V_{H}(\phi_{H})
\end{equation}

where, $\dot{\theta}=\frac{Q}{a^{3}\phi_{H}^{2}}$.\\
Now, taking $T(\phi_{D})=m \dot{\phi}_{D}^{2}$ and scale factor
$a(t)=t^{n}$; we get from the Einstein field equations (3) and (4)
that

\begin{equation}
\frac{n}{t^{2}}=\frac{1}{2}\left(\dot{\phi}_{H}^{2}-\frac{Q^{2}}{t^{6n}\phi_{H}^{2}}+m
\dot{\phi}_{D}^{2} \right)
\end{equation}

Choosing
$\dot{\phi}_{D}^{2}=\frac{2n}{m}\left(\frac{1}{t^{2}}\right)$ we
get

\begin{equation}
\phi_{H}=\sqrt{\frac{2Q}{3n-1}}t^{1-3n}
\end{equation}

\begin{equation}
\phi_{D}=\sqrt{\frac{2n}{m}}\ln t
\end{equation}

Using the conservation equations (14) and (15) with $p_{X}\equiv
p_{H}$ and $\rho_{X}\equiv \rho_{H}$ we get, under interaction,
the forms of the potentials as

\begin{equation}
V_{H}=\frac{1}{4}t^{-3n\delta}\left[(3n-1)Qt^{3n(-2+\delta)}\left(\frac{4\delta}{2-\delta}+
\frac{t^{-2+6n}(2-3n(2+\delta))}{3n\delta-2}\right)+C_{1}\right]
\end{equation}

and

\begin{equation}
V_{D}=\frac{2n^{2}\left(2+3n\delta+\sqrt{\frac{n}{n-1}}(1-3n\delta)\right)}{mt^{2}(2+3n\delta)}+C_{2}t^{3n\delta}
\end{equation}

\begin{figure}
\includegraphics[height=1.8in]{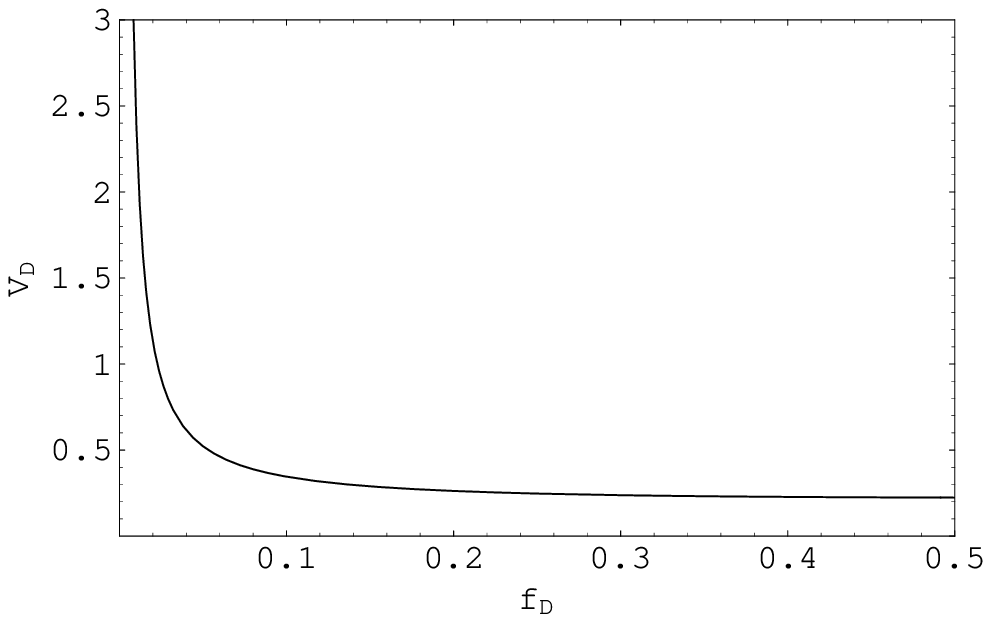}~~~~~
\includegraphics[height=1.8in]{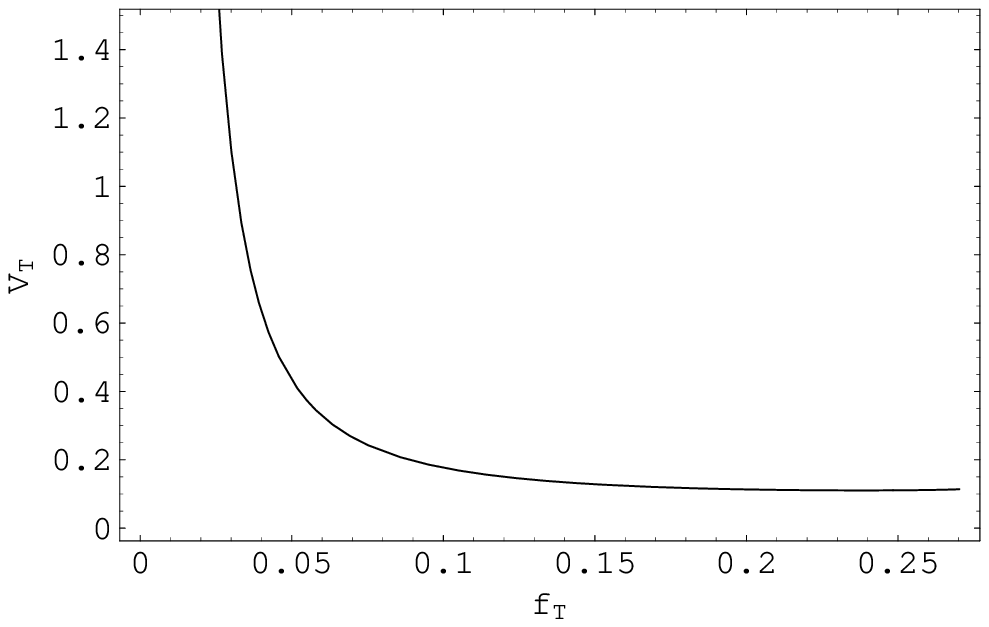}~\\
\vspace{1mm} ~~~~~~~~~~~~Fig.8~~~~~~~~~~~~~~~~~~~~~~~~~~~~~~~~~~~~~~~~~~~~~~~~~~~~~~Fig.9\\
\vspace{6mm} Figs. 8 and 9 show the variation of $V_{D}$ and
$V_{T}$ against $\phi_{D}$ and $\phi_{T}$  in the case of
interaction between DBI-essence and tachyonic field.

\vspace{6mm} \vspace{6mm}
\end{figure}

\begin{figure}
\includegraphics[height=1.8in]{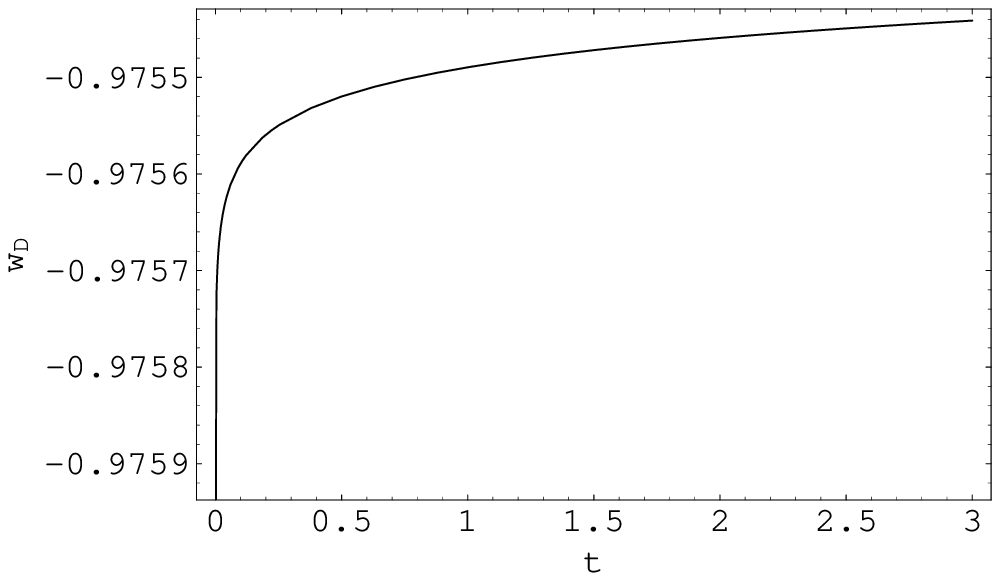}~~~~~
\includegraphics[height=1.8in]{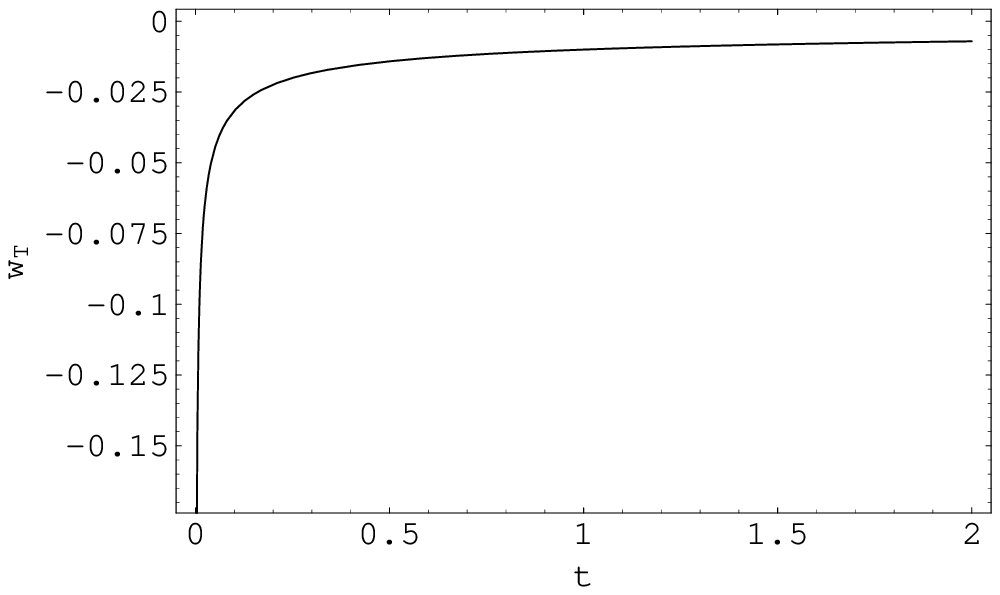}~\\
\vspace{1mm} ~~~~~~~~~~~~Fig.10~~~~~~~~~~~~~~~~~~~~~~~~~~~~~~~~~~~~~~~~~~~~~~~~~Fig.11\\
\includegraphics[height=1.8in]{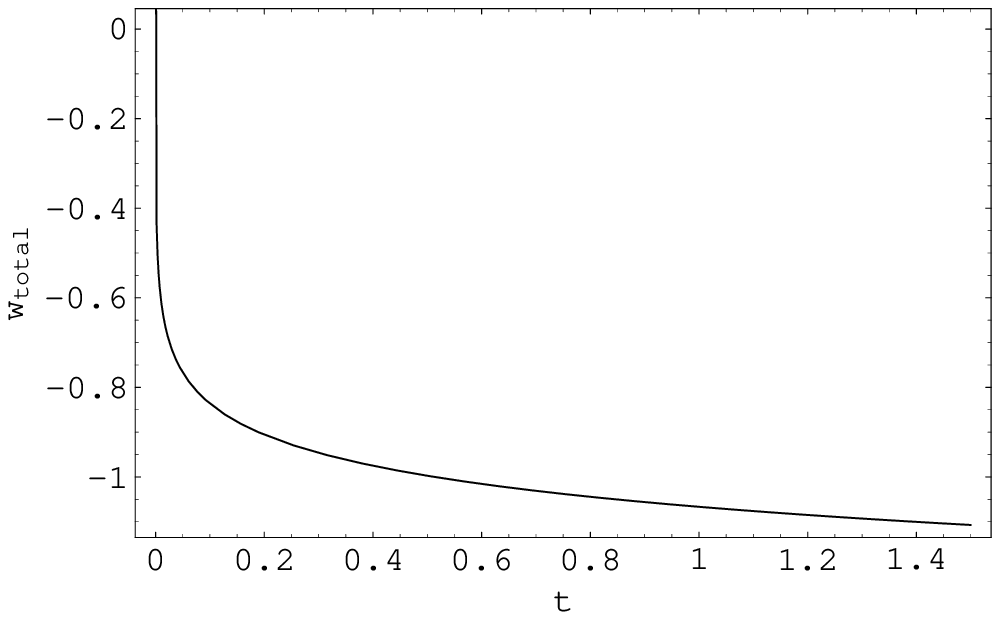}~\\
\vspace{1mm} ~~~~~~~~~~~~Fig.12\\
\vspace{6mm} Figs. 10 and 11 show the evolution of
$\omega_{D}=\frac{p_{D}}{\rho_{D}}$ and
$\omega_{T}=\frac{p_{T}}{\rho_{T}}$ with $t$. in the situation of
interaction between tachyonic field and DBI-essence.\\
Fig. 12 shows the evolution of
$\omega_{total}=\frac{(p_{T}+p_{D})}{(\rho_{T}+\rho_{D})}$ with
$t$ in the said interaction.

\vspace{6mm} \vspace{6mm}
\end{figure}

\section{\normalsize\bf{Interaction with tachyonic field}}

The action for the homogeneous tachyon condensate of string theory
in a gravitational background is given by [16]

\begin{equation}
S=\int\sqrt{-g}d^{4}x\left[\frac{\cal R}{16\pi G}+\cal L\right]
\end{equation}

where, $\cal L$ is the Lagrangian density given by

\begin{equation}
{\cal L} = -V(\phi)\sqrt{1+
g^{\mu\nu}\partial_{\mu}\phi\partial_{\nu}\phi}
\end{equation}

where $\phi$ is the tachyonic field, $V(\phi)$ is the tachyonic
field potential and $\cal R$ is the Ricci Scalar. Since we are
considering the interacting situation, we would denote the
potential for the tachyonic field by $V_{T}$ and the corresponding
scalar field by $\phi_{T}$. The energy density $\rho_{T}$ and
pressure $p_{T}$ for tachyonic field are given by

\begin{equation}
\rho_{T}=\frac{V_{T}(\phi_{T})}{\sqrt{1-\dot{\phi}_{T}^{2}}}
\end{equation}

\begin{equation}
p_{T}=-V_{T}(\phi_{T})\sqrt{1-\dot{\phi}_{T}^{2}}
\end{equation}

In the conservation equations (14) and (15) we consider
$\rho_{X}\equiv \rho_{T}$ and $p_{X}\equiv p_{T}$ and assume that
$1-\dot{\phi}_{T}^{2}=t^{n}$ and scale factor $a=t^{m}$. Then from
the conservation equation (15) we get

\begin{equation}
V_{T}=e^{\frac{3nt^{-m}}{m}}t^{-m+3n(1+\delta)}
\end{equation}

Using the above form of tachyonic field potential under
interaction in the field equations (8) and (9) with $T=k_{1}
\dot{\phi}_{D}^{2}$ and $\dot{\phi}_{D}^{2}=t^{k_{2}}$ we get the
form of potential of DBI-essence under interaction with tachyonic
field as

\begin{equation}
V_{D}=\frac{1}{2}\left(\frac{6n}{t^{2}}+4k_{1}t^{k_{2}}+(3-4k_{1})\sqrt{\frac{k_{1}}{k_{1}-1}}t^{k_{2}}+
\frac{e^{\frac{3nt^{-m}}{m}}t^{-m+3n(1+\delta)}(3+t^{m})}{\sqrt{t^{m}}}\right)
\end{equation}

From equations (11) and (12)we get

\begin{equation}
\omega_{D}=\frac{(k_{1}-\sqrt{k_{1}(k_{1}-1)})t^{k_{2}}-V_{D}(\phi_{D})}{k_{1}\left(\sqrt{\frac{k_{1}}{k_{1}-1}}-1
\right)t^{k_{2}}+V_{D}(\phi_{D})}
\end{equation}

which can be simplified by taking $V_{D}$ from equation (47).
Furthermore, from equations (45) and (46) it can be easily
obtained that

\begin{equation}
\omega_{T}=-t^{n}
\end{equation}

Evolutions of $\omega_{D}$ and $\omega_{T}$ have been presented in
figures 10 and 11 respectively.
\\

\section{\normalsize\bf{Interaction with new agegraphic dark energy}}
 Quantum mechanics together with general relativity leads to the
  Karolyhazy relation and a corresponding energy density of quantum fluctuations of
space-time. Based on this energy density, Cai [17] proposed a dark
energy model, the so-called agegraphic dark energy model, in which
the age of the universe is introduced as the length measure. The
corresponding energy density is given by [17]

\begin{equation}
\rho_{q}=\frac{3n^{2}m_{p}^{2}}{T^{2}}
\end{equation}

where,

\begin{equation}
T=\int\frac{da}{Ha}
\end{equation}

\begin{figure}
\includegraphics[height=1.8in]{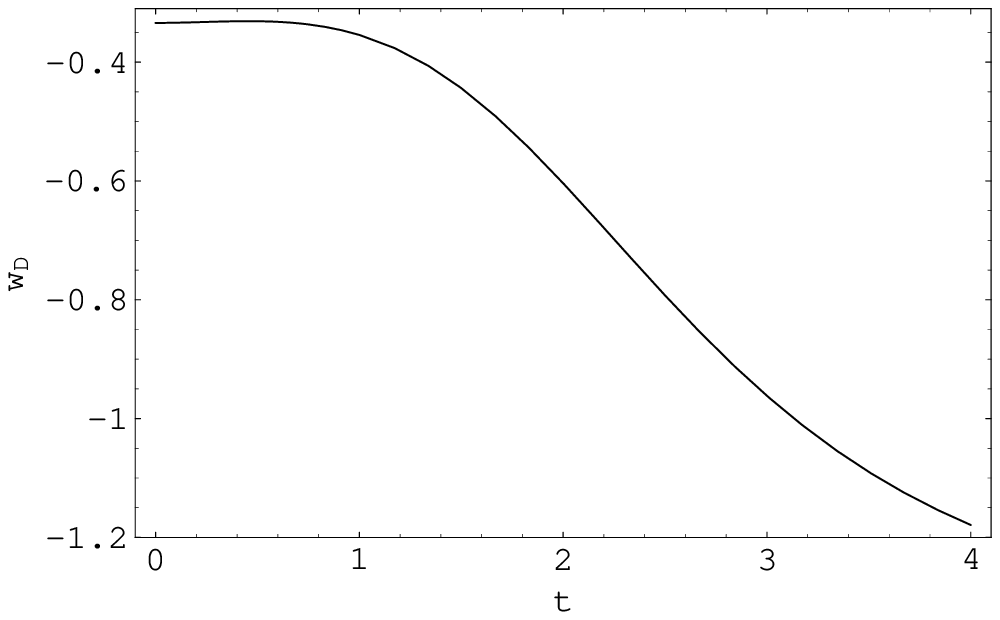}~~~~~~
\includegraphics[height=1.8in]{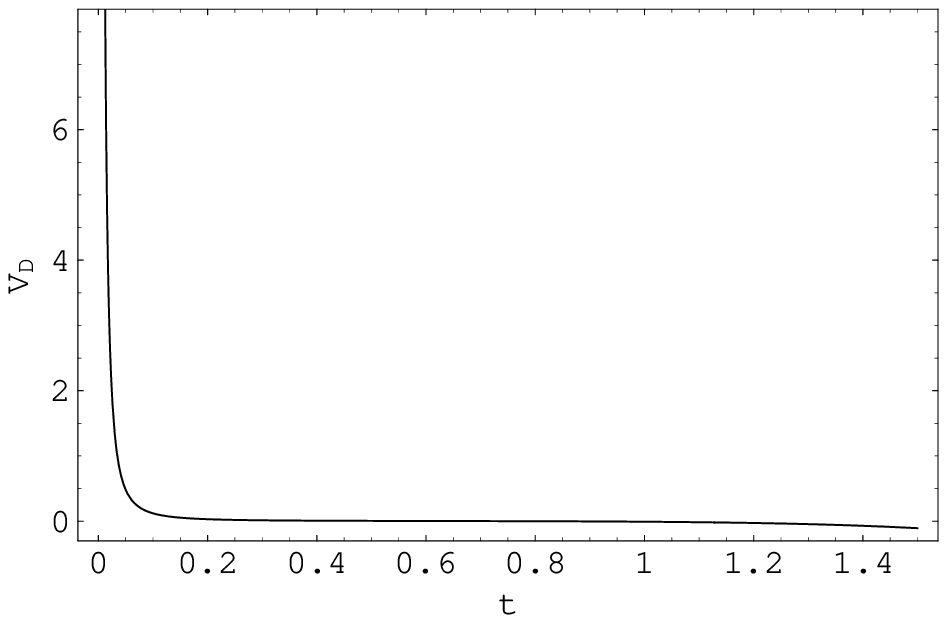}~\\
\vspace{1mm} ~~~~~~~~~~~~Fig.13~~~~~~~~~~~~~~~~~~~~~~~~~~~~~~~~~~~~~~~~~~~~~~~~~~~Fig.14\\
\vspace{6mm} Figs. 13 and 14 show the variation of $\omega_{D}$
and $V_{D}$ against $t$ when DBI-essence is interacting with new
agegraphic dark energy.

\vspace{6mm} \vspace{6mm}
\end{figure}

If we consider a flat FRW universe containing the agegraphic dark
energy and pressureless-matter, the the corresponding Friedman
equation becomes

\begin{equation}
H^{2}=\frac{1}{3m_{p}^{2}}(\rho_{m}+\rho_{q})
\end{equation}

Introducing the fractional energy densities
$\Omega_{i}=\rho_{i}/3m_{p}^{2}H^{2}$ we can get

\begin{equation}
\Omega_{q}=\frac{n^{2}}{H^{2}T^{2}}
\end{equation}

and the equation-of-state parameter $\omega_{q}=p_{q}/\rho_{q}$ is
given by

\begin{equation}
\omega_{q}=-1+\frac{2}{3n}\sqrt{\Omega_{q}}
\end{equation}

 The agegraphic dark energy was constrained by using some old high
redshift objects and type Ia supernovae [18]. A new agegraphic
dark energy model was proposed in [19], where the time scale is
chosen as the conformal time $\eta$ instead of the age of the
universe. For this new agegraphic dark energy, the energy density
$\rho_{A}$ is given as

\begin{equation}
\rho_{A}=\frac{3n^{2}m_{p}^{2}}{\eta^{2}}
\end{equation}

where

\begin{equation}
\eta=\int\frac{dt}{a}
\end{equation}

Thus, $\dot{\eta}=1/a$. The corresponding fractional energy
density is given by

\begin{equation}
\Omega_{A}=\frac{n^{2}}{H^{2}\eta^{2}}
\end{equation}

In the present work, we would consider an interaction between
DBI-essence and new agegraphic dark energy. To do so, we consider
equations (14) and (15) and in the present case $\rho_{X}\equiv
\rho_{A}$ and $p_{X}\equiv p_{A}$. Considering the interaction we
obtain the equation-of-state parameter as

\begin{equation}
\omega_{A}=-1-\delta+\frac{2}{3n}\frac{\sqrt{\Omega_{A}}}{a}
\end{equation}

Now we assume $a=t^{k_{1}}$, $1-\dot{\phi}_{D}^{2}=t^{k_{2}}$, and
$T=k_{3}\dot{\phi}_{D}^{2}$ and get the potential of DBI-essence
under interaction as

\begin{equation}
V_{D}=\frac{1}{2}\left(\frac{6k_{1}}{t^{2}}+(1-t^{k_{2}})\left(4k_{3}-\sqrt{\frac{k_{3}}{k_{3}-1}}(4k_{3}-3)\right)
-\frac{3}{k_{1}}(k_{1}-1)^{2}(-2+3k_{1}(2+\delta))n^{2}m_{p}^{2}t^{2(k_{1}-1)}\right)
\end{equation}

Using equations (11) and (12), we get the equation of state for
DBI-essence under this interaction as

\begin{equation}
\omega_{D}=\frac{(k_{3}-\sqrt{k_{3}(k_{3}-1)})(1-t^{k_{2}})-V_{D}(\phi_{D})}{k_{3}\left(\sqrt{\frac{k_{3}}{k_{3}-1}}-1
\right)(1-t^{k_{2}})+V_{D}(\phi_{D})}
\end{equation}

which can be simplified by taking $V_{D}$ from equation (59).
Evolution of this equation of state parameter $\omega_{D}$ under
the given interaction has been presented in figure 13. From this
figure, we find that the equation of state parameter is decreasing
with time and always negative during evolution.\\

\begin{figure}
\includegraphics[height=1.8in]{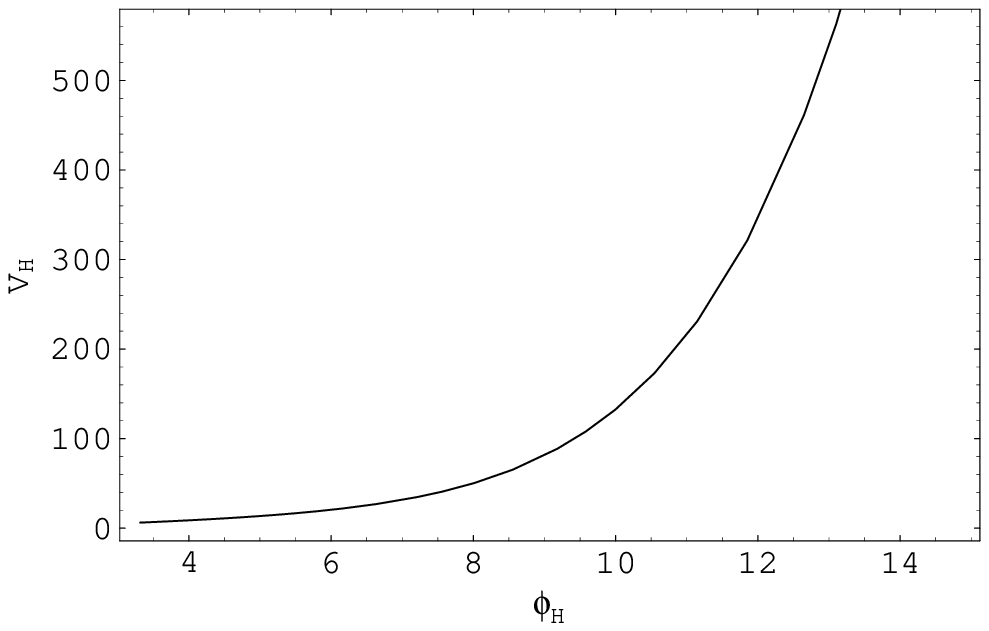}~~~~~~
\includegraphics[height=1.8in]{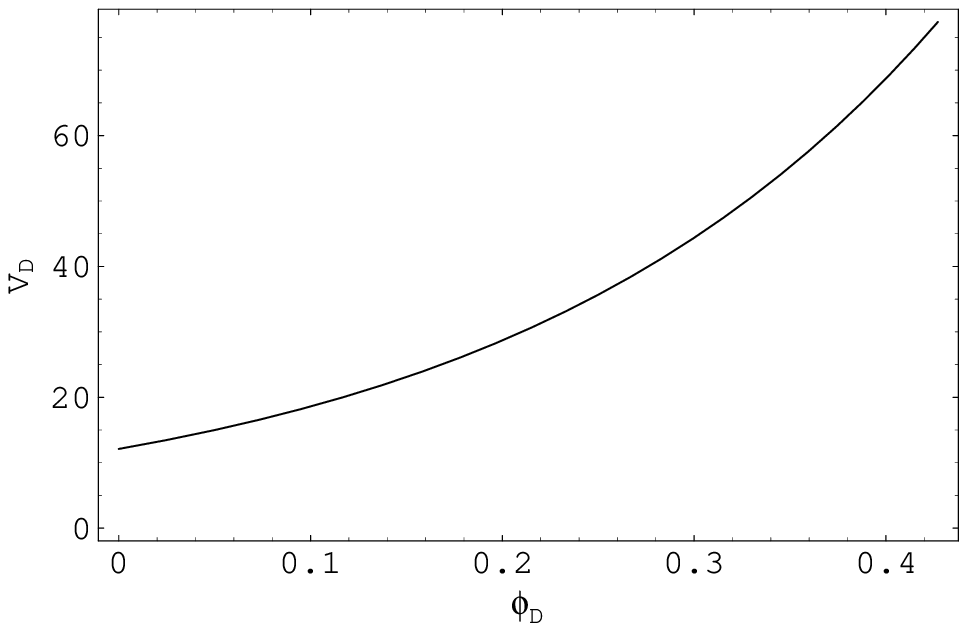}~\\
\vspace{1mm} ~~~~~~~~~~~~Fig.15~~~~~~~~~~~~~~~~~~~~~~~~~~~~~~~~~~~~~~~~~~~~~~~~~~~Fig.16\\
\includegraphics[height=1.8in]{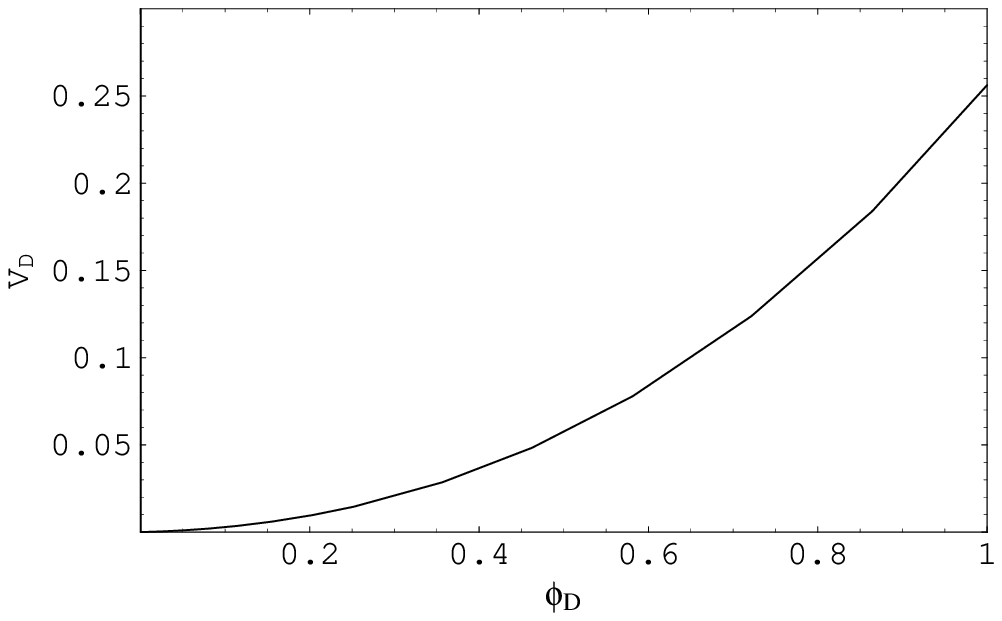}~~~~~~~
\includegraphics[height=1.8in]{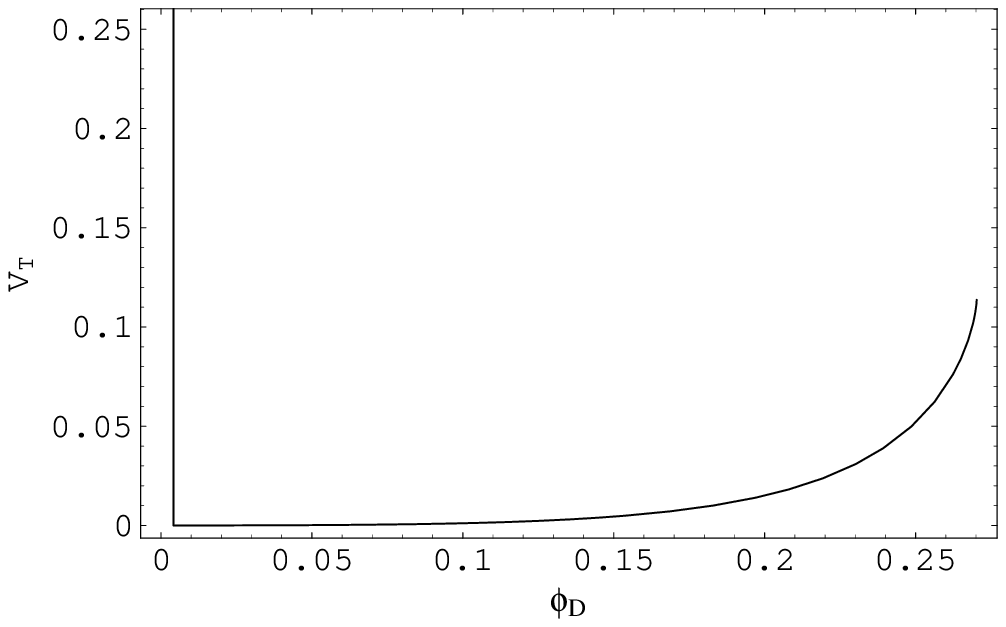}~\\
\vspace{1mm} ~~~~~~~~~~~~Fig.17~~~~~~~~~~~~~~~~~~~~~~~~~~~~~~~~~~~~~~~~~~~~~~~~~~Fig.18\\
\vspace{6mm} Figs. 15 and 16 show the variation of potential with
scalar field when DBI-essence is interacting with hessence with
$\delta=0.6$.\\Figs. 17 and 18 show the variation of potential
with scalar field when DBI-essence is interacting with tachyonic
field with $\delta=0.6$.

\vspace{6mm} \vspace{6mm}
\end{figure}

\section{\normalsize\bf{Conclusions}}

In the present work, we have considered the DBI-essence as a
candidate of dark energy interacting with other candidates of dark
energies. The other candidates considered here are the modified
Chaplygin gas, tachyonic field, hessence and new agegraphic dark
energy. We have considered an interaction parameter $\delta$ and
accordingly generated the conservation equations in presence of
interaction. First we have taken the modified Chaplygin gas as a
dark energy interacting with DBI-essence. In this case we have
derived the energy density of the modified Chaplygin gas in the
presence of interaction, i.e. involving the interaction parameter
$\delta$. The scalar field for the DBI-essence has been derived in
the presence of interaction and its evolution with time has been
viewed in figure 1, which shows that the scalar field is
increasing with time. From figure 2, which shows the evolution of
the potential of the DBI-essence in this interacting situation,
reveals that there is a sharp decrease in the potential with
increase in the scalar field. Next we have considered the
interaction between the DBI-essence and hessence. In figures 3, 4,
and 7 we have seen that the potentials for the DBI-essence and
hessence are falling rapidly with increase in time. In figures 5
and 6, the evolution of potentials are viewed with increase in the
scalar field. In those figures it is revealed that the potentials
are falling with increase in the scalar fields in the situation of
interaction between DBI-essence and hessence. However, the fall is
not as sharp as it is in the case of the evolution of the
potential with increase in time. The next interaction is between
the DBI-essence and tachyonic field. In figures 8 and 9 we have
seen that the potentials for both of the dark energy candidates
are decreasing with increase in the corresponding scalar fields.
In this interaction, we have also considered the equation of the
individual as well as the combination of the two interacting dark
energies in the figures 10, 11 and 12. In all the cases we find
that the equation of state parameters are negative in sign and in
the case of combination of the two dark energies it is falling
sharply and finally getting stable in the vicinity of $-1$.
Whereas, in the individual cases, the parameter is increasing with
increase in time. Finally, we have considered the interaction
between DBI-essence and new agegraphic dark energy. In figure 13,
we have plotted the equation of state parameter for DBI-essence in
this interacting situation against time $t$. We find in this
figure that the said parameter is gradually decreasing with time
and in lying approximately between $-1.2$ and $0$. The potential
in this case is showing the same sharply falling pattern with
increase in $t$ as it was in the previous interactions. It is,
therefore, observed from the interactions that the potentials are
always experiencing sharp fall with increase in time in all the
four cases. However, the fall in the potential is not so sharp
when it is considered against the corresponding scalar field. It
should be noted that in all the cases we have taken the
interaction parameter $\delta$ very close to 0. Similar nature in
the figures have been observed in the case when the interaction
parameter is negative. Now we examine the $V(\phi)$-$\phi$
relations if the interaction term is taken at larger value. In the
case of interaction with modified Chaplygin gas, $\delta$ was
taken equal to 0.005. Taking larger values we find that the
similar nature as it was earlier is available. In the case of
interaction with hessence, we had taken $\delta=0.006$. Now we
take $\delta=0.6$, and we find that the potentials are increasing
with increase in the scalar fields (figures 15 and 16). Now we
consider the interaction with tachyon. Earlier we had taken
$\delta=0.05$. Now, taking $\delta=0.6$ we find that the potential
$V$ is increasing with increase in the scalar field $\phi$
(figures 17 and 18).
\\

{\bf Acknowledgement:}\\

The authors are thankful to IUCAA, Pune, India for warm
hospitality where part of the work was carried out. Also UD is
thankful to UGC, Govt. of India for providing research project grant (No. 32-157/2006(SR)).\\

{\bf References:}\\
\\
$[1]$  E. J. Copeland, M. Sami, and S. Tsujikwa, {\it Int. J. Mod.
Phys. D} {\bf 15} 1753 (2006); S. Perlmutter et al., {\it
Astrophys. J.} {\bf 517} 565 (1999).\\
$[2]$ A. G. Riess et al., {\it Astrophys. J.} {\bf 116} 1009
(1998); A. G. Riess et al., {\it Astron. J.} {\bf 117} 707 (1999);
S. Boughn and R. Crittenden,
{\it Nature} {\bf 427} 45 (2004)\\
$[3]$ J. Martin and M. Yamaguchi, {\it Phys. Rev. D} {\bf 77}
123508 (2008); L. P. Chimento , R. Lazkoz,  and I. Sendra, {\it
Gen. Relativ. Gravit.} DOI:10.1007/s10714-009-0901-z (2009). \\
$[4]$ S. Weinberg, {\it Rev. Mod. Phys.} {\bf 61} 1 (1989); D. N.
Spergel et al. [arXiv:astro-ph/0603449].\\
$[5]$ L. P. Chimento, R. Lazkoz, and I. Sendra,
[arXiv:0904.1114v2].\\
$[6]$ S. Kachru et al., {\it JCAP} {\bf 10} 013 (2003); E.
Silverstein and D. Tong, {\it Phys. Rev. D}
{\bf 70} 103505 (2004).\\
$[7]$ B. Gumjudpai and J. Ward, {\it Phys. Rev. D} {\bf 80} 023528
(2009); J. Martin and M. Yamaguchi, {\it Phys. Rev. D} {\bf 77}
123508 (2008).\\
$[8]$ J. E. Lidsey and I. Huston, {\it JCAP} {\bf 0707} 002
(2007); D. Baumann and L. McAllister, {\it Phys. Rev. D} {\bf 75}
123508 (2007).\\
$[9]$ M. S. Berger, H. Shojaei, {\it Phys. Rev. D} {\bf 74} 043530
(2006).\\
$[10]$ R. Herrera, D. Pavon and W. Zimdahl, {\it Gen. Rel. Grav.}
{\bf 36} 2161 (2004); R. G. Cai and A. Wang, {\it JCAP} {\bf 0503}
002 (2005); Z. K. Guo, R. G. Cai and Y. Z. Zhang, {\it JCAP}
{\bf 0505} 002 (2005); T. Gonzalez and I. Quiros, [gr-qc/0707.2089].\\
$[11]$ H. Zhang and Z. Zhu, {\it Phys. Rev. D} {\bf 73} 043518
(2006); U. Debnath, {\it Astrophys. Space Sci.} {\bf 312} 295 (2007).\\
$[12]$  U. Alam,  V. Sahni, T. D. Saini and A. A. Starobinsky,
{\it Mon. Not. R. Astron. Soc.} {\bf 344} 1057 (2003); V. Gorini,
 A. Kamenshchik and U. Moschella, {\it Phys. Rev. D} {\bf 67} 063509
 (2003); M. C. Bento, O. Bertolami and A. A. Sen, {\it Phys. Rev. D} {\bf 66} 043507
(2002).\\
$[13]$ T. Barreiro and A. A. Sen, {\it Phys. Rev. D} {\bf 70}
124013 (2004).\\
$[14]$ H. Wei, R. G. Cai and D. F. Zeng, {\it Class. Quantum
Grav.} {\bf 22} 3189 (2005); Z. K. Guo, Y. S. Piao, X. M. Zhang
and Y. Z. Zhang, {\it Phys. Lett. B} {\bf 608} 177 (2005); X. F.
Zhang, H. Li, Y. S. Piao and X. M. Zhang, {\it Mod. Phys. Lett. A}
{\bf 21} 231 (2006); H. Wei, N. Tang, and S. N. Zhang, {\it Phys.
Rev. D} {\bf 75} 043009 (2007),W. Zhao, {\it Phys. Lett. B} {\bf 655} 97 (2007).\\
$[15]$ H. Wei, R. G. Cai, {\it Phys. Rev. D} {\bf 72} 123507
(2005) ; M. Alimohammadi, H. M. Sadjadi, {\it Phys. Rev. D} {\bf
73} 083527 (2006); H. Wei, N. Tang and S. N. Zhang, {\it Phys.
Rev. D} {\bf 75} 043009 (2007).\\
$[16]$ S. Chattopadhyay, U. Debnath and G. Chattopadhyay, {\it
Astrophys. Space Sci} {\bf 314 } 41 (2008).\\
$[17]$  R. G. Cai, {\it Phys. Lett. B} {\bf 657} 228 (2007);
J. Zhang, X. Zhang and H. Liu, {\it Eur. Phys. J. C.}{\bf 54} 303 (2008).\\
$[18]$H. Wei and R. G. Cai, {\it Phys. Lett. B} {\bf 663} 1(2008);
Y. Zhang, H. Li, X. Wu, H. Wei and R. G. Cai, arXiv:0708.1214.\\
$[19]$  H. Wei and R. G. Cai, {\it Phys. Lett. B} {\bf 660} 113
(2008).\\

\end{document}